\documentstyle[amssymb,aps,prb,preprint]{revtex}

\begin{document}
\title{Hall Effect of ${\rm {\bf La_{2/3}(Ca,Pb)_{1/3}MnO_3}}$ Single Crystals near
the Critical Temperature}
\author{S. H. Chun, M. B. Salamon, and P. D. Han}
\address{Department of Physics and Materials Research Laboratory, University of\\
Illinois at Urbana-Champaign, Urbana, Illinois 61801-3080}
\date{\today}
\maketitle

\begin{abstract}
The Hall resistivity $\rho _{xy}$ of a La$_{2/3}$(Ca,Pb)$_{1/3}$MnO$_3$
single crystal has been measured as a function of temperature and field. The
overall behavior is similar to that observed previously in thin-films. At 5
K, $\rho _{xy}$ is positive and linear in field, indicating that the
anomalous contribution $R_S$ is negligible. However, the effective carrier
density in a free electron model is $n_{eff}=2.4$ holes/Mn, even larger than
the 0.85-1.9 holes/Mn reported for thin-films and far larger than the 0.33
holes/Mn expected from the doping level. As temperature increases, a strong,
negative contribution to $\rho _{xy}$ appears, that we ascribe to $R_S$.
Using detailed magnetization data, we separate the ordinary $(\propto B)$
and anomalous $(\propto M)$ contributions. Below $T_C$, $\left| R_S\right|
\propto \rho _{xx},$ indicating that magnetic skew scattering is the
dominant mechanism in the metallic ferromagnetic regime. At and above the
resistivity-peak temperature, we find that $\rho _{xy}/\rho _{xx}M$ is a
constant, independent of temperature and field. This implies that the
anomalous Hall coefficient is proportional to the magnetoresistance. A
different explanation based on two fluid model is also presented.
\end{abstract}


The paramagnetic insulator to ferromagnetic metal transition of the
manganite system La$_{1-x}$A$_x$MnO$_3$ (where A is Ca, Sr, or Pb) has long
been thought to be roughly described by the double exchange interaction\cite
{Zener}. After the recent rediscovery of colossal magnetoresistance in this
system\cite{Jin}, a renewed effort was made to understand the physical
properties\cite{Ramirez}. It now appears that double exchange alone cannot
explain the large change in conductivity, and that Jahn-Teller effects
should be included\cite{Millis}. This view is supported by the evidence of
strong coupling between lattice effects, magnetic order, and transport
behavior\cite{Hwang}.

A consequence of the Jahn-Teller/double-exchange picture is that the
charge-carrier characteristics change from a fully spin polarized
(half-metallic) band at low temperatures, through a regime of partially
polarized bands, before finally becoming localized as polarons well above $%
T_C$. It is important, therefore, to explore the band properties of the
three regimes by means of the Hall effect. The situation is complicated by
the presence of the extraordinary Hall coefficient in the ferromagnetic
phase. We report here the first complete measurement of both the ordinary
and extraordinary Hall coefficients on single crystal samples spanning the
full temperature range, from metallic ferromagnet to polaronic conductor. As
in previous measurements on thin films and low temperature measurements on
single crystals, we find the ordinary Hall coefficient to be unexpectedly
small at low temperature, corresponding to a hole concentration of 2.4
holes/Mn atom in a simple one-carrier free-electron model. Other researchers
have reported values between 0.85 and 1.9 holes/Mn\cite
{Snyder,Matl,Jacob,Asamitsu}. We show that this result can be reconciled
with the actual doping level when details of the Fermi surface are taken
into account.

In order to extract the extraordinary Hall coefficient, and to treat the
data in the presence of ``colossal'' magnetoresistance near $T_C$, we have
combined detailed $\rho _{xy}$ measurements with magnetization ($M$) and
longitudinal resistivity ($\rho _{xx}$) measurements on the same La$_{2/3}$%
(Ca,Pb)$_{1/3}$MnO$_3$ single crystal. While it is possible to separate
ordinary and extraordinary effects at low temperature, only an electron-like
transverse resistivity is found in the magnetoresistive regime near $T_C$.
In this temperature range, the tangent of the Hall angle, $\rho _{xy}/\rho
_{xy}$ is found be remarkably linear in the measured magnetization, despite
300 \% changes in the longitudinal resistivity. As it seems unlikely that
the skew-scattering mechanism continues to operate in the hopping regime, we
consider another approach in which the observed conductivity is assumed to
be a field- and temperature-dependent admixture of band-like and polaronic
charge carriers, conducting in parallel\cite{TF}. This approach also
provides a very good representation of the data, but the polaronic Hall
constant that emerges increases more rapidly as the temperature approaches $%
T_C$ than is expected from conventional polaron theory.

High quality single crystals of La$_{2/3}$(Ca,Pb)$_{1/3}$MnO$_3$ were grown
from 50/50 PbF$_2$/PbO flux. Specimens were cut along crystalline axes from
larger pre-oriented crystals. Contact pads for Hall resistivity measurements
were made by Au evaporation and standard photolithographic patterning,
followed by ion-milling. Au wires and silver paint were used for electrical
connections. The contact resistances after annealing were less than 1 $%
\Omega $ at room temperature. We adopted a low frequency ac method for the
measurements. First, the transverse voltage signal was nulled at zero field
by a potentiometer at each temperature. The change in the transverse
resistance as a function of field was amplified by SR552 low noise
preamplifier, and recorded by a phase sensitive detector. Magnetization was
measured by a 7 T SQUID magnetometer.

The crystal has a fairly small residual resistivity ($\rho _{xx}^0\approx $
51 $\mu \Omega $cm) and large magnetoresistance (326 \% at 293 K under 7 T).
The metal-insulator transition temperature, determined by the maximum change
in resistivity $d\rho _{xx}/dT$ under zero magnetic field, is 287.5 K.
Figure 1 shows the field dependence of $\rho _{xy}$ at several temperatures.
The overall behavior is the same as those of thin film samples\cite
{Matl,Jacob}. Far below $T_C$, $\rho _{xy}$ decreases at first and then
increases, showing a hole-like high field Hall coefficient. The initial drop
becomes larger as $T$ approaches $T_C$. Around $T_C$, $\rho _{xy}$ is
strongly curved making a simple interpretation impossible. Far above $T_C$, $%
\rho _{xy}$ shows a negative Hall coefficient, characteristic of
electron-like charge carriers.

In ferromagnetic metals, the embedded magnetic moments cause asymmetric
scattering of current-carrying electrons, which in turn produce an
additional transverse voltage, called anomalous (or extraordinary) Hall
effects. The anomalous Hall field is proportional to the current density and
the sample magnetization, so $\rho _{xy}$ is generally written as\cite{Hurd} 
\begin{equation}
\rho _{xy}\left( B,T\right) =R_H\left( T\right) B+\mu _0R_S\left( T\right)
M\left( B,T\right) ,  \label{1}
\end{equation}
where $R_S\left( T\right) $ is the temperature dependent anomalous Hall
coefficient. From separate magnetization measurements corrected for
demagnetizing fields, we could extract $R_H$ and $R_S$ from $\rho _{xy}$ for
temperatures below $T_C,$ as shown in the lower inset of Fig. 2. If we
assume the free electron model for the Hall coefficient, the effective
charge carrier density, $n_{eff}=1/eR_H,$ turns out to be 4.1$\times $10$%
^{22}$ cm$^{-3}$ or 2.4 holes/Mn below 100 K, a value much larger than the
nominal doping level (0.33 holes/Mn), so the naive interpretation is not
valid here. Others report similar results and some authors interpreted this
as an effect of charge carrier compensation\cite{Matl,Jacob}. We will do the
same, as follows. Pickett and Singh calculated the $T=0$ band structure of
1/3-doped manganites, and concluded that the alloy is nearly half-metallic,
and that the majority-spin band consists of a spherical Fermi surface
containing 0.05 electrons and a nearly cubic Fermi surface containing 0.55
holes, larger than the nominal doping level of 0.33\cite{Pickett}. Since our
Hall experiment as well as others is in the weak field limit even at the
lowest temperature and at the highest field ($\omega _c\tau \sim 0.01\ll 1$,
where $\omega _c$ is the cyclotron frequency and $\tau $ the Drude
relaxation time), the Hall coefficient $R_H$ of a non-spherical Fermi
surface is given not by the high field limit $R_\infty =1/ne$, but rather by 
$R_H=r/ne$, where $r$ is a dimensionless factor depending on the details of
the Fermi surface. For example, $r$ is known to equal 1/2 for a cubic Fermi
surface\cite{GAD}. In a two band model, the Hall coefficient is given by $%
R_H=(r_hn_h\mu _h^2-r_en_e\mu _e^2)/e(n_h\mu _h+n_e\mu _e)^2$. If we assume
equal mobilities for holes and electrons, we obtain $n_{eff}V_c=V_c/eR_H=1.6$
holes/Mn, where $V_c$ is the volume per formula unit. This explains why the
observed $R_H$ is much smaller than expected from the doping level. Jacob 
{\it et al}. assumed $\mu _e/\mu _h=2.1$ to have a quantitative agreement to
their results without considering the shape dependent factor $r$\cite{Jacob}%
. They justified their assumption based on the density of states at the
Fermi level. Using the electron and hole densities from Pickett and Singh,
we can reproduce the low temperature value of $R_H$ by assuming a mobility
ratio $\mu _e/\mu _h=1.5$.

The high field slope of $\rho _{xy}$ changes with temperature, becoming
steeper as $T$ approaches $T_C$ from below, and changing sign above $T_C$ to
show an electron-like Hall coefficient. Because the anomalous Hall effect,
which tends to saturate at high fields, cannot explain this slope change, we
have to assume a temperature dependent $n_{eff}$. The main panel of Fig. 2
shows the temperature dependence of $n_{eff}$ in the free electron
approximation with $r=1$. As $T$ increases, $n_{eff}$ decreases slowly until
250 K and then drops rapidly. At and above $T_C$, Eq. (1) cannot decompose $%
\rho _{xy}(B)$ because of the polaronic contribution and large
magnetoresistance which will be discussed later. A qualitatively similar
change was reported by Jacob {\it et al}.\cite{Jacob}. Although the physical
origin is not certain, the participation of minority spin band in the charge
transport, which is evidenced by decreasing polarization as $T$ increases,
could be one of the reasons. We compared $n_{eff}(T)$ with $\rho _{xx}(T)$
or $M(T)$, but no simple relation was found. In particular, no logarithmic
dependence on $M(T)$ was seen\cite{Booth}.

The decomposition shown in the lower inset of Fig. 2 enables us to extract
the temperature dependence of anomalous Hall coefficient $R_S$ also. It is
customary in ferromagnetic metals to compare $R_S$ with $\rho _{xx}$ to
determine the origin of the anomalous Hall effects. From 10 K to 270 K,
where $R_S$ is well-defined, we find $R_S$ is proportional to $\rho _{xx}$
(Fig. 2, the upper inset). The coefficient $\alpha $ is $-1.7\times 10^{-3}$
T$^{-1}$ and the absolute magnitude of $R_S$ is comparable to that of
Ca-doped thin film samples\cite{Matl,Jacob}. This linear relation is in
agreement with the classical skew scattering theory, where moving charge
carriers experience a force due to the magnetic field produced by a
localized magnetic moment and are scattered asymmetrically\cite{Maranzana}.
However, the sign difference relative to the ordinary Hall coefficient and
the absence of an $R_S$ peak below $T_C$ disagree with the predictions. We
note here that Kim {\it et al}. presented an interesting explanation for the
opposite sign between $R_H$ and $R_S$ from the appearance of topological flux%
\cite{Kim}. The presence of $R_S$ and its proportionality to $\rho _{xx}$
supports our earlier argument that the metallic resistivity is dominated by
spin-dependent scattering\cite{Jaime2}.

Around and above $T_C,$ $\rho _{xy}(B)$ is not easy to interpret because of
the large magnetoresistance in this region and the electron-like polaronic
contribution as verified by Jaime {\it et al}.\cite{Jaime}. A strong
curvature in $\rho _{xy}(B)$ was observed at temperatures where the
zero-field $\rho _{xx}$ shows a peak and the magnetoresistance is the
largest (the inset of Fig. 3). Considering the large changes in both $\rho
_{xy}$ and $\rho _{xx}$ as a function of field, it is instructive to
consider the Hall angle, defined by $\tan \theta _H=\rho _{xy}/\rho _{xx}$.
Interestingly, if we plot $\tan \theta _H$ as a function of the sample
magnetization, all the data above 310 K fall on a single line crossing zero
as shown in Fig. 3. This means that we can describe the Hall resistivity in
this temperature regime simply as 
\begin{eqnarray}
\rho _{xy}\left( B,T\right) &=&\alpha ^{\prime }\mu _0M\left( B,T\right)
\rho _{xx}\left( B,T\right)  \label{2} \\
&=&\mu _0R_S^{\prime }\left( B,T\right) M\left( B,T\right) ,  \nonumber
\end{eqnarray}
where the field and temperature dependent anomalous Hall coefficient $%
R_S^{\prime }\left( B,T\right) =\alpha ^{\prime }\rho _{xx}\left( B,T\right) 
$ and $\alpha ^{\prime }=-1.4\times 10^{-3}$ T$^{-1}$. The high temperature
coefficient $\alpha ^{\prime }$ is a little smaller than the low temperature
one $\alpha $. The linear relationship between the Hall angle and the
magnetization was already noted by Matl {\it et al}., but only in the low
field regime\cite{Matl}. Our results show that the Hall angle scales with
the magnetization over the entire field investigated for temperatures at and
above the resistivity peak temperature. In Fig. 4, we compare $\rho _{xy}$
data with the calculation from Eq. (2) as a function of field (thin solid
lines). The agreement is quite good except for some high field data at 320
K. The origin of $R_S^{\prime }$ is not certain at this time. The skew
scattering theory, which was applied to low temperature data, assumes an
interaction between band electrons and localized magnetic moments. The
current understanding is, however, that the charge carriers are dominantly
small polarons and the conduction is by means of hopping processes at
temperatures where Eq. (2) holds. Furthermore, the universal relation
between the Hall angle and magnetization despite more than 300 \% change in
magnetoresistance requires a more general theory that does not depend on the
scattering length of charge carriers.

For high enough temperature, the adiabatic small polaron hopping theory
predicts the Hall mobility, $\mu _H\equiv \tan \theta _H/B$, is an
increasing function of temperature\cite{Emin}, while the present result is
the opposite since $\mu _H\varpropto M/B=\chi $. In manganite system, the
high temperature Hall effect results are in accord with the small polaron
theory\cite{Jaime}. So Eq. (2) might be valid for limited temperature regime
and should change to activated form as temperature increases. In order to
have an explanation with wider application range, we have tried another
method motivated by a phenomenological two fluid model, recently proposed by
one of the authors\cite{TF}. They assume the charge carriers in manganites
move through two parallel channels: one has metallic conductivity $\sigma
_m(T)$, and the other has hopping conductivity $\sigma _h(T)$. The key
element of this model is the introduction of the {\it mixing factor} $%
c\left( B,T\right) $, which is the portion of band electrons in total
carriers and absorbs the field dependence of the conductivity. Then, the
total conductivity reads as 
\begin{equation}
\sigma _{tot}(B,T)=c\sigma _m(T)+\left( 1-c\right) \sigma _h(T),  \label{3}
\end{equation}
and $\sigma _m(T)$ and $\sigma _h(T)$ are assumed to be field independent,
and are extracted by fitting the conductivity data far from $T_C$. The
mixing factor is calculated from Eq. (3) by assuming that $\sigma _m(T)$ and 
$\sigma _h(T)$ can be extrapolated to the region nearer $T_C$. This two
fluid model can be applied to the Hall effect in a similar way as in
multiband model as following: 
\begin{equation}
\frac{\rho _{xy}}B=\frac{R_mc^2\sigma _m^2+R_h(1-c)^2\sigma _h^2}{(c\sigma
_m+(1-c)\sigma _h)^2},  \label{4}
\end{equation}
where $R_m=(R_1+\alpha \mu _0\sigma _m^{-1}M/B)/c$ and $R_h=R_2/(1-c)$ are
effective Hall coefficients due to band electrons and polarons,
respectively. $R_1$ and $R_2$ are free parameters for fitting. As we can see
in Fig. 4, the agreement with the data is excellent (thick dashed lines).
The normal Hall coefficient of metallic phase $R_1$ corresponds to about 1
holes/Mn and slightly temperature dependent. The resulting values of $R_2$,
the polaron Hall coefficient, are a rapidly decreasing function of
temperature, which is more or less proportional to the magnetic
susceptibility $\chi $. When Emin and Holstein's theory valid for polarons
in a non-magnetic material is applied to the temperature dependence of $R_2$%
\cite{Emin}, we obtain $\Delta =4500$ meV, five times larger than that
determined by the conductivity curve. Since Emin and Holstein's theory was
derived for high temperature limits, it may not apply close to the
metal-insulator transition. It is quite likely that these polarons have
significant magnetic character and that the double exchange mechanism
continues to strongly affect their mobility near $T_C$.

In conclusion, we measured $\rho _{xy}$, $\rho _{xx}$, and $M$ of a La$%
_{2/3} $(Ca,Pb)$_{1/3}$MnO$_3$ single crystal and observed a change from
hole-like $R_H$ below $T_C$ to electron-like $R_H$ far above $T_C$. We
obtained 2.4 holes/Mn at 5 K and interpreted as a result of carrier
compensation. We also found a linear relation between the negative anomalous
Hall coefficient and zero field $\rho _{xx}$ below $T_C$ in accord with the
magnetic skew scattering theory. At and above the resistivity peak
temperature, we found that $\rho _{xy}/\rho _{xx}M$ is a constant,
independent of temperature and field. This implies that the anomalous Hall
coefficient is proportional to the magnetoresistance. Another interpretation
based on recently proposed two fluid model also produces a good agreement
with the data, but the temperature dependence of polaronic contribution to
the Hall effect is different from the high-temperature-limit prediction.

This work was supported in part by DOE DEFG-91ER45439.

\begin{figure}[tbp]
\caption{The Hall resistivity $\rho_{xy}$ of a ${\rm %
La_{2/3}(Ca,Pb)_{1/3}MnO_3}$ single crystal as a function of field at
indicated temperatures.}
\label{fig1}
\end{figure}

\begin{figure}[tbp]
\caption{The effective number of holes/Mn as a function of temperature. The
lower inset shows the decomposition of $\rho_{xy}$ into ordinary and
anomalous Hall effects below $T_C$. The upper inset shows the linear
relation between the anomalous Hall coefficient $R_S$ and the longitudinal
resistivity $\rho_{xx}$.}
\label{fig2}
\end{figure}

\begin{figure}[tbp]
\caption{The scaling behavior of the Hall angle, $\tan \theta_H=\rho_{xy} /
\rho_{xx}$ and the sample magnetization $M$ at temperatures from 310 K to
350 K. The inset shows the colossal magnetoresistance of this material.
Solid circles indicate the region of scaling.}
\label{fig3}
\end{figure}

\begin{figure}[tbp]
\caption{The measured Hall resistivity data as a function of field at
selected temperatures above $T_C$. Thin solid lines come from the assumption
of field and temperature dependent $R_S$. Thick dashed lines are from two
fluid model. see text for details.}
\label{fig4}
\end{figure}

\end{document}